\documentclass[journal]{IEEEtran}
\usepackage{amsmath}
\usepackage{amssymb}
\usepackage{amsfonts}
\usepackage{graphicx}
\usepackage{epsfig}
\usepackage{subfigure}
\usepackage{psfrag}
\usepackage{cite}
\usepackage{latexsym}
\usepackage{url}
\usepackage{color}
\usepackage{multirow}
\usepackage{bm}
\usepackage{multicol,comment}
\usepackage{stfloats}

\usepackage[colorlinks,linkcolor=black,anchorcolor=black,citecolor=black]{hyperref}
\begin{document}
\title{Online Maneuver Design for UAV-Enabled NOMA Systems via Reinforcement Learning}
\author{Yuwei Huang$^1$, Xiaopeng Mo$^{2,3}$, Jie Xu$^{3,4}$, Ling Qiu$^1$, and Yong Zeng$^{4,5}$\\
$^1$School of Information Science and Technology, University of Science and Technology of China\\
$^2$School of Information Engineering, Guangdong University of Technology\\
$^3$Future Network of Intelligence Institute (FNii) and the School of Science and Engineering, The Chinese University of Hong Kong, Shenzhen\\
$^4$National Mobile Communications Research Laboratory, Southeast University\\
$^5$The Purple Mountain Laboratory, Nanjing \\
E-mail:~hyw1023@mail.ustc.edu.cn,~xiaopengmo@mail2.gdut.edu.cn,\\~xujie@cuhk.edu.cn,~lqiu@ustc.edu.cn,~yong\_zeng@seu.edu.cn}
\maketitle

\newcommand{\mv}[1]{\mbox{\boldmath{$ #1 $}}}
\newtheorem{lemma}{\underline{Lemma}}[section]
\newtheorem{remark}{\underline{Remark}}[section]
\newtheorem{proposition}{\underline{Proposition}}[section]
\newtheorem{assumption}{\underline{Assumption}}[section]
\begin{abstract}
This paper considers an unmanned aerial vehicle (UAV)-enabled uplink non-orthogonal multiple-access (NOMA) system, where multiple users on the ground send independent messages to a UAV via NOMA transmission. We aim to design the UAV's dynamic maneuver in real time for maximizing the sum-rate throughput of all ground users over a finite time horizon. Different from conventional offline designs considering static user locations under deterministic or stochastic channel models, we consider a more challenging scenario with mobile users and segmented channel models, where the UAV only causally knows the users' (moving) locations and channel state information (CSI). Under this setup, we first propose a new approach for UAV dynamic maneuver design based on reinforcement learning (RL) via Q-learning. Next, in order to further speed up the convergence and increase the throughput, we present an enhanced RL-based approach by additionally exploiting expert knowledge of well-established wireless channel models to initialize the Q-table values. Numerical results show that our proposed RL-based and enhanced RL-based approaches significantly improve the sum-rate throughput, and the enhanced RL-based approach considerably speeds up the learning process owing to the proposed Q-table initialization.
\end{abstract}
\section{Introduction}
Unmanned aerial vehicle (UAV)-enabled wireless communications have been envisioned as one of the important technologies for the fifth-generation (5G) and beyond wireless networks\cite{RZhangUAVchallenges}, where UAVs are employed as aerial wireless platforms (such as relays \cite{RZhangrelay} and base stations (BSs)\cite{JXubroadcast,QWuscp}) to serve end users on the ground for providing data access, enhancing coverage, and improving communication rates. In particular, owing to the superiority in providing higher spectral efficiency and supporting massive connectivity of non-orthogonal multiple-access (NOMA) technique\cite{NOMA}, UAV-enabled NOMA systems have been regarded as a promising candidate on the road to integrating UAVs into 5G and beyond networks. Substantial research efforts have been devoted to this research paradigm. For example, \cite{NOMA} jointly designed the trajectory and transmit power of the UAV to serve ground NOMA users and highlighted the design challenges of integrating NOMA into UAV networks. Furthermore, by considering capacity-achieving NOMA transmission strategies, the authors in \cite{JXubroadcast} and \cite{peimingNOMA} characterized the fundamental rate limits of UAV-enabled broadcast channel \cite{JXubroadcast} and UAV-enabled multiple access channel (MAC) \cite{peimingNOMA}, respectively, by optimizing the UAV's trajectory jointly with wireless resource allocations.

In the UAV communications literatures, there have been a growing number of prior works investigating the UAV's placement optimization \cite{staticaltitude,statichorizontal,peiming} and trajectory optimization \cite{RZhangrelay,JXubroadcast,QWuscp,NOMA,peimingNOMA,saad,lifeng} for both scenarios with quasi-stationary and flying UAVs, respectively. On one hand, when the UAV is placed at a static location over the whole communication period of interest, \cite{staticaltitude} optimized the UAV's altitude to maximize the coverage probability on the ground, while \cite{statichorizontal,peiming} optimized the locations of multiple UAV-BSs to maximize the minimum throughput of all ground users. On the other hand, under the flying UAV scenario, \cite{RZhangrelay} jointly designed the UAV-relay's trajectory and power allocation to maximize the end-to-end throughput. In \cite{JXubroadcast,QWuscp,NOMA,peimingNOMA,saad}, the UAV was employed as aerial BS to serve ground users, for which its trajectory together with the resource allocation (e.g., power allocation and user association) were jointly optimized to enhance the system performance. In addition, when the UAVs are employed as aerial users with their own missions other than dedicated to supporting wireless communications, the authors in \cite{shuowen} designed the UAV's trajectory to minimize the complete mission time, under the connection constraints with ground BSs. Besides the optimization based designs, in recent years, reinforcement learning (RL) has found abundant applications in wireless communication networks \cite{machinelearning}, particularly  in UAV communications (see, e.g.,\cite{RL1,RL2,RL3,RL4}). Generally speaking, most of these works adopted RL to solve an offline optimization problem, in which the UAV trajectory was designed by training it over a finite time horizon repeatedly under the stationary environment.

Despite the recent research progress, existing works have assumed that the UAV knows the ground users' {\it static} locations, and can predict the channel state information (CSI) over time based on the assumption that the UAV-to-ground channels follow deterministic channels (e.g., line-of-sight (LoS) channel)\cite{statichorizontal,peiming,RZhangrelay,JXubroadcast,QWuscp} or stochastic channels (e.g., probabilistic LoS channel) \cite{peimingNOMA,saad}. Then, based on the assumptions on user locations and CSI, the UAV placement/trajectory design can be formulated as deterministic optimization problems, which are solvable offline via convex/non-convex optimization techniques or RL. Nevertheless, in practice, the ground users may move randomly over time, as such the UAV cannot know their locations over time prior to communication. Besides, the channels in practical scenarios are more complicated than LoS or probabilistic LoS models due to the uneven distribution of obstacles (e.g., buildings and trees), so that it is difficult to acquire the accurate modeling and the perfect information of all relevant parameters. As a result, it is very difficult, if not impossible, for the UAV to acquire the CSI over time prior to communication. Thus, how to design the UAV's dynamic maneuver to track the mobile users' movement to provide better service in UAV-enabled NOMA systems with CSI uncertainty and under more practical channel models is a challenging problem that is not well addressed yet.

In particular, this paper studies a UAV-enabled uplink NOMA system, where multiple users on the ground send independent messages to a UAV via NOMA transmission. Our objective is to maximize the sum-rate throughput of all ground users over a finite time horizon by designing the UAV's dynamic maneuver, subject to practical constraints on the UAV's initial location and flight speed. Different from prior offline designs considering static user locations under LoS\cite{statichorizontal,peiming,RZhangrelay,JXubroadcast,QWuscp}  or probabilistic LoS channel models \cite{peimingNOMA,saad}, we consider a more challenging scenario with segmented channel models and mobile ground users. It is assumed that the UAV only causally knows the users' locations and CSI. As a result, this problem becomes challenging to solve. To tackle this problem,  we adopt the RL, a powerful technique of machine learning \cite{RL}, which can solve the sequential decision problems without requiring the full information of the environment. First, we propose a new RL-based approach to design the UAV dynamic maneuver by utilizing Q-learning. Then, we propose a novel enhanced RL-based approach to speed up the convergence by exploiting the well-established wireless channel models as expert knowledge to help initialize Q-table values. Finally, numerical results show that our proposed RL-based and enhanced RL-based approaches achieve significant throughput gains over the heuristic design based on average channel models, and the enhanced RL-based approach considerably improves the convergence speed. 

It is worth noting that although RL has been widely used to design the UAV maneuver for navigation and obstacles avoidance (see, e.g., \cite{UAVcontrol}), which ignored the communication issue considered in this paper. Besides, this paper is the first attempt proposing a novel enhanced RL-based approach for UAV dynamic maneuver design by exploiting the well-established wireless channel models as expert information to help initialize the Q-table to further speed up the convergence.

\section{System Model}
In this paper, we consider the uplink NOMA transmission in a UAV-enabled wireless communication system, in which a set of mobile users on the ground, denoted as $\mathcal{K}\triangleq\{1,...,K\}$,  send independent messages to a UAV flying in the sky. We focus on a finite period $\mathcal {T}=[0,T]$ with duration $T>0$, which are discretized into $N$ time slots each with identical duration $\delta=T/N$. Here, $\delta$ is chosen to be sufficiently small such that the UAV's location can be assumed to be approximately constant within each time slot even at its maximum flying speed. We denote $\mathcal{N}\triangleq\{1,...,N\}$.

We consider a three-dimensional (3D) Cartesian coordinate system, where each user $k$ has a zero altitude and moves slowly over time. Let $\mv w_{k}[n]=(x_{k}[n],y_{k}[n])$ denote the horizontal location of user $k$ at slot $n\in\mathcal N$. The UAV flies at a fixed altitude $H>0$ in meter (m) with the time-varying horizontal location being $\mv q[n]=(x[n],y[n])$ at slot $n\in\mathcal N$, in which $H$ is chosen to be based on the practical constraint on the minimum UAV flying altitude imposed by government regulations for safety consideration\cite{shuowen}. Suppose that the UAV's initial horizontal location is pre-determined as $\hat{\mv q}_{I}=(x_{I},y_{I})$, and thus we have $\mv q[1]=\hat{\mv q}_{I}$. At each slot $n\in\mathcal N$, the UAV can change its location or hover at the same location. For ease of analysis, we define $\tilde{\mathcal{A}}=\{(0,0),(-\lambda,0),(\lambda,0),(0,\lambda),(0,-\lambda)\}$ as the set of UAV's possible location changes over one slot, each element of which corresponds to hovering, flying left, right, forwards, and backwards, respectively. Here, $\lambda$ is a constant denoting the UAV's displacement at each slot. Let $\mv\lambda[n]$ denote the UAV's location change at each slot $n$. We thus have $\mv \lambda[n]\in\tilde{\mathcal A}$ and $\mv q[n]=\mv q[n-1]+\mv{\lambda}[n-1],~\forall n\in\mathcal N\backslash\{1\}$.

Different from conventional designs based on LoS or probabilistic LoS channels, in this paper, we consider a more practical segmented channel model as in \cite{radiomap2}, in which if there are obstacles between the UAV and the user $k$, the channel follows a non-line-of-sight (NLoS) model, and otherwise, the channel follows a LoS model. Thus, the channel power gain from the user $k$ to the UAV is denoted by $h_{k,b}[n]=\mu_{k,b}[n]\xi_{k,b}[n]\beta_{k,b}[n]d_{k}[n]^{-\alpha_{k,b}[n]}$, where $\alpha_{k,b}[n]$ is the path loss exponent, $\beta_{k,b}[n]$ is the reference channel power gain, $\xi_{k,b}[n]$ denotes the shadowing component, $\mu_{k,b}[n]$ accounts the small-scale fading, and $b\in\{\text{LoS,NLoS}\}$ represents the strong dependence of the propagation parameters on LoS or NLoS scenario. In the following, we drop the subscript $b$ for ease of exposition.

We consider the uplink NOMA transmission over the MAC from the $K$ users to the UAV. At slot $n\in\mathcal N$, let $u_{k}[n]$ denote the message (with unit power) transmitted by each user $k\in\mathcal K$. Accordingly, the received signal at the UAV is given by $\small{\tilde{s}[n]=\sum_{k\in\mathcal K}\sqrt{\tilde{P}h_{k}[n]}u_{k}[n]+v[n]}$, where $\tilde{P}$ is the fixed transmit power of user $k$, and $v[n]$ denotes the additive white Gaussian noise (AWGN) at the UAV receiver with power $\sigma^{2}$. To achieve the capacity region of MAC, the UAV adopts the successive interference cancellation (SIC) to decode the information \cite{wireless}, and the $K$ users can employ Gaussian signaling with $u_{k}[n]$'s being independent circularly symmetric complex Gaussian (CSCG) random variables with zero mean and unit variance. Specifically, at each time slot $n$, denote $\mv\varphi=[\varphi(1),...,\varphi(K)]$ as the decoding order at the UAV, which indicates that the UAV receiver first decodes the message $u_{\varphi(K)}[n]$ from user $\varphi(K)$, then decodes $u_{\varphi(K-1)}[n]$ by cancelling the interference from $u_{\varphi(K)}[n]$, followed by $u_{\varphi(K-2)}[n]$,$u_{\varphi(K-3)}[n]$, and so on, until $u_{\varphi(1)}[n]$. Therefore, under any given $\mv \varphi$, the achievable rate by user $\varphi(k)$ at each slot $n$ in bits/second/Hertz (bps/Hz) is given by $$\small{r_{\varphi(k)}[n]=\log_{2}\left(\frac{\sum\limits_{j=1}^{k}\tilde{P}h_{\varphi(j)}[n]+\sigma^{2}}{\sum\limits_{i=1}^{k-1}\tilde{P}h_{\varphi(i)}[n]+\sigma^{2}}\right),~\forall k\in\mathcal K.}$$ As a result, the sum-rate throughput of the $K$ users at each slot $n$ is expressed as follows, which is irrelevant to the decoding order $\mv \varphi$\cite{wireless}.
\begin{align}
\tilde{R}[n]=\sum_{k\in\mathcal K}r_{\varphi(k)}[n]=\log_{2}(1+\tilde{P}\sum_{k\in\mathcal K}h_{k}[n]/\sigma^{2}).\label{rate}
\end{align}

Our objective is to maximize the average sum-rate throughput of all ground users\footnote{How to ensure the individual data rates of different users is another interesting problem for UAV-enabled NOMA communications. Towards this end, it is very crucial to additionally design the decoding order and transmit power control at these users, which, however, is left for our future work.}over the duration consisting of $N$ time slots (i.e., $\frac{1}{N}\sum_{n=1}^{N}\tilde{R}[n]$), by optimizing the UAV's maneuver $\{\mv q[n]\}$. Thus, the problem of our interest is formulated as
\begin{align}
\text{(P1):}\max\limits_{\{\bm q[n],\bm\lambda[n]\in\tilde{\mathcal A}\}}~&\frac{1}{N}\sum\limits_{n=1}^{N}\log_{2}(1+\tilde{P}\sum\limits_{k\in\mathcal K}h_{k}[n]/\sigma^{2})\nonumber\\
\text{s.t.}~&\mv q[1]=\hat{\mv q}_{I},\label{initialconstraint}\\
&\mv q[n]=\mv q[n-1]+\mv{\lambda}[n-1],\forall n\in\mathcal N\backslash\{1\},\label{speedconstraint}
\end{align}
where (\ref{initialconstraint}) and (\ref{speedconstraint}) denote UAV's initial location and maneuver constraints, respectively. It is assumed that to achieve the data-rate throughput at each slot $n$, the UAV is able to know the users' locations and CSI at the current time slot by using proper localization and channel estimation algorithms; however, the UAV does not know the users' locations and CSI for future slots due to their random mobility. Thus, prior to communication, the UAV is not aware of the CSI $\{h_{k}[n]\}$ over time, and accordingly does not know the objective function of (P1). In this case, problem (P1) is challenging to solve.

\section{Proposed Solution to (P1) via RL}
In this section, we first propose a novel approach to solve (P1) based on RL by using Q-learning, and then propose an enhanced RL-based approach to help speed up the convergence and further improve the throughput, in which the wireless models are utilized as expert knowledge to help initialize the values of Q-table.

\subsection{RL-based Approach for UAV Maneuver Design}
Before proceeding, we briefly introduce RL and Q-learning. RL is an area of machine learning, which studies how an agent can take actions in an environment to maximize the cumulative reward over a certain time horizon \cite{RL}. Q-learning is a specific RL method, which enables the agent to find an optimal policy for reward maximization, via updating a Q-table that returns the expected rewards under different actions.

In particular, we describe the Q-learning based on a Markov decision process (MDP) $\{\mathcal S,\mathcal A,P,R\}$ with state space $\mathcal S$, action space $\mathcal A$, state transition probability $\small{P_{a}(s,s')=Pr(s_{n+1}=s'|s_{n}=s,a_{n}=a)}$, and reward function $R_{a}(s,s')$. At each step $n$, the agent observes its current state $s_{n}$, selects and performs an action $a_{n}\in\mathcal A$, then moves into the subsequent state $s_{n+1}$ and receives an immediate reward $r_{n}=R_{a_{n}}(s_{n},s_{n+1})$. In general, the goal of the agent is to determine a policy $\pi$ to maximize the received reward, where the policy is defined as  $\pi(a|s)=Pr(a_{n}=a|s_{n}=s)$. Then, define the Q-function (or state-action value function) as the expected discounted reward for executing action $a$ at state $s$ based on $\pi$. Thus, for policy $\pi$, the Q-function is given by
\begin{align}
Q^{\pi}(s,a)=\mathbb{E}_{\pi}\{ R_{n}|s_{n}=s,a_{n}=a\},\label{Q-function}
\end{align}
where $\mathbb{E}\{\cdot\}$ denotes the statistic expectation, and
\begin{align}
R_{n}=\sum_{i=0}^{N-1}\gamma^{i}R_{n+i+1}\label{reward}
\end{align}denotes the discounted summation of all future rewards at current step $n$, with $\gamma\in[0,1)$ denoting the discount factor to balance the tradeoff between long-term and short-term rewards. Particularly, $\gamma\to 1$ represents that the agent focuses on the long-term reward, while $\gamma\to 0$ denotes that the agent only considers the immediate reward. Accordingly, the agent's objective is to find an optimal policy to maximize the Q-function, i.e., $\pi^{*}(a|s)=\text{arg max}_{a\in\mathcal A}Q^{\pi*}(s,a)$.\footnote{In practical implementation, the agent may need to act randomly with a certain probability for exploring unknown states, as explained in detail later.} Notice that if the optimal policy $\pi^{*}$ is not unique, then we can simply choose any one of these policies without loss of optimality. By combining (\ref{Q-function}) and (\ref{reward}), the Q-table can be updated in each step $n$ based on the Bellman update equation as
\begin{align}
&Q^{\pi}(s_{n},a_{n})=Q^{\pi}(s_{n},a_{n})\nonumber\\
&+\alpha(r_{n}+\gamma\max_{a}Q^{\pi}(s_{n+1},a)-Q^{\pi}(s_{n},a_{n})),\label{update}
\end{align}
where $\alpha\in[0,1]$ denotes the learning rate to determine how much the old information is retained. To ensure the convergence of Q-learning, the learning parameters $\alpha$ and $\gamma$ should be chosen properly.

Now, we explain how to use RL to solve (P1) with implicit objective functions based on Q-learning. Specifically, we use a Q-table to approximate the Q-functions, and set the state space as $\mathcal S=\{\mv q[n]\}$ for each slot $n$ and the action space as $\mathcal A=\tilde{\mathcal A}$.\footnote{Notice that in order to implement the table-based Q-learning to design the UAV's dynamic maneuver, the number of states should be finite. Towards this end, we divide the whole area into finite grids, which will be discussed in detail in Section $\text{\uppercase\expandafter{\romannumeral4}}$.} Then, the RL-based approach proceeds as follows.

Consider any one particular slot $n\in\mathcal N$, where the UAV stays at location $\mv q[n]$ with current state $s_{n}=\mv q[n]$. In this case, the UAV chooses an action (location change) $a_{n}=\mv{\lambda}[n]\in\mathcal A$ by employing the $\epsilon$-greedy policy based on the Q-table, such that the UAV chooses a random action within $\mathcal A$ with probability $\epsilon\in[0,1]$, and chooses the action that maximizes the Q-function, i.e., $a=\text{arg max}_{\tilde{a}\in\mathcal A}Q(s_{n},\tilde{a})$, with probability $1-\epsilon$. Based on this action, the UAV then moves to a new location $\mv q[n+1]=\mv q[n]+\mv\lambda[n]$ (corresponding to the subsequent state $s_{n+1}$), and receives an immediate reward as the instantaneous sum-rate throughput $\tilde{R}[n]$ in (\ref{rate}), i.e., $r_{n}=R_{a_{n}}(s_{n},s_{n+1})=\tilde{R}[n]$. Next, the current Q-value, i.e., $Q^{\pi}(s_{n},a_{n})$, is updated according to (\ref{update}). The above procedure proceeds until $n=N$. By properly choosing the parameter $\epsilon$, the UAV can efficiently balance the trade-off between exploring unvisited locations (states) to get more experiences versus exploiting already tried locations to gain higher rewards, which is referred to as the {\it exploration-exploitation trade-off} in RL literature \cite{RL}. In general, $\epsilon$ should be chosen to be exponentially decreasing over time.  Furthermore, in our considered dynamic environment with mobile users, the discount factor should be chosen to focus on the long-term reward, i.e., $\gamma\to1$.

The practical implementation of Q-learning critically depends on the initial Q-table at slot $n=1$. In our proposed RL-based approach, the initial Q-table values are set to zero similar as \cite{RL1}. However, it generally takes a long time to converge towards the optimal solution, which may considerably compromise the overall throughput performance during the whole communication period of interest.

\subsection{Enhanced RL-based Approach for UAV Maneuver Design}
To tackle the slow convergence issue of the RL-based design, in this subsection we propose a novel enhanced RL-based approach by exploiting wireless expert information to aid the Q-learning. In the following, we first explain how to roughly predict the CSI and estimate rate functions by using expert information, and then propose a procedure to properly initialize the Q-table based on such estimations.

In the wireless communications society, there have been various well-established models on wireless channels and communication rates, which have been overlooked in the above proposed RL-based approach with zero Q-table initially. Based on this observation, we propose to exploit such knowledges to help initialize the Q-table. More specifically, based on the rate function in (\ref{rate}) under NOMA transmission, it is evident that we can (roughly) estimate the communication rate over time based on only coarse channel information. Though we do not know the exact channel environment and CSI, we know that any wireless channels generally follow a simplified stochastic model with path loss, shadowing, and small-scale fading \cite{wireless}. In addition, for UAV-to-ground communications, we additionally know that the channel power gain is dependent on the elevation angle, which can be characterized by the LoS probability in an average/stochastic sense.\footnote{Notice that the probabilistic  LoS model can only reveal the stochastic property of UAV-to-ground channels over a certain area. For any specific link, the wireless channel can either be LoS or NLoS.} In this case, we can use the average channel power gain as a rough channel prediction, i.e.,\cite{probability}

\vspace{-10pt}\begin{small}\begin{align}
\bar{h}_{k}[n]=p_{k,\text{LoS}}[n]\bar{\beta} d_{k}[n]^{-\bar{\alpha}}+\eta (1-p_{k,\text{LoS}}[n])\bar{\beta}d_{k}[n]^{-\bar{\alpha}},\label{plos}
\end{align}\end{small}where $p_{k,\text{LoS}}[n]=1/(1+C\exp(-D(\text{arcsin}(H/d_{k}[n])-C)))$ is the LoS probability for the link between the UAV and user $k$ with $C$ and $D$ being the probability parameters, $\bar{\beta}$ is the average reference channel power gain over all user-UAV links, $\bar{\alpha}$ is the average path loss exponent over all user-UAV links, and $\eta<1$ is an additional attenuation factor due to the NLoS propagation. Note that although the parameters $C$, $D$, $\eta$, $\bar{\beta}$, and $\bar{\alpha}$ are not known by the UAV, their values are limited within certain ranges based on empiric results (as explained in Section $\text{\uppercase\expandafter{\romannumeral4}}$). Thus, we properly choose these parameters for getting rough channel predictions. In the special case with LoS UAV-to-ground channels, we have $\small{p_{k,\text{LoS}}[n]=1,~\forall k\in\mathcal K,n\in\mathcal N}$, which becomes an accurate channel model (instead of stochastic) and is widely used in UAV maneuver design \cite{statichorizontal,RZhangrelay}.

Based on such rough channel prediction models, we estimate the initial values of Q-table by adding additional training procedure, which is implemented similarly as that in the proposed RL-based approach in Section $\text{\uppercase\expandafter{\romannumeral3}}$-A. The only difference is that during the training procedure, the UAV computes the rewards by using the above rough channel estimations in (\ref{plos}) based on the users' initial locations (as we cannot predict their movement). After convergence, the training procedure obtains a Q-table that will be used as the initial one for the proposed RL-based approach in Section $\text{\uppercase\expandafter{\romannumeral3}}$-A. If the adopted ``average'' channel model perfectly matches with the real environment (with static users), the obtained Q-table is generally optimal. Even if it does not match with the time-varying environment in practical scenarios with mobile users, the obtained-initial Q-table can still give rough information about the achievable rewards (throughputs) at different UAV locations, thus helping speed up the convergence of the subsequent Q-learning, as will be shown next in simulations.

Notice that due to the employment of the additional training procedure, our proposed enhanced RL-based approach generally has a higher computation complexity than the proposed RL-based approach with zero Q-table initially. Despite this fact, as the training procedure can be implemented offline at the UAV prior to communication, the enhanced RL-based approach is feasible in practical implementation, and there generally exists a fundamental tradeoff between the training time versus the training accuracy.

\section{Numerical Results}
\begin{figure}
\centering
\includegraphics[width=7cm]{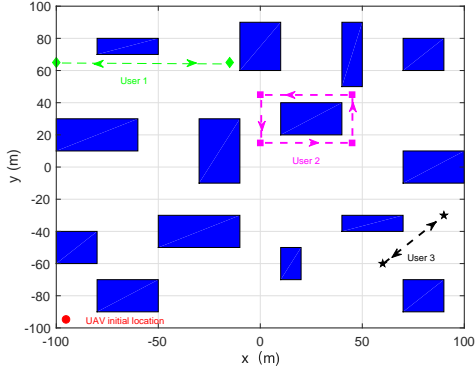}
\caption{Simulation setup with blue cubes denoting obstacles (e.g., buildings).}
\label{environment}
\end{figure}
In this section, we present numerical results to validate the performance of our proposed RL-based and enhanced RL-based designs. Unless otherwise stated, we set the users' transmit power as $P=\text{23 dBm}$, the UAV's fixed flight altitude as $H=\text{100 m}$, and the noise power as $\sigma^{2}=\text{--80 dBm}$. Furthermore, we consider a square area of size $200~\text{m}~\times~200~\text{m}$ with $K=3$ mobile users\footnote{Notice that we consider $K=3$ and assume that the users move based on the pre-determined trajectories in the numerical results for ease of exposition. Our proposed solution is applicable for any number of ground users and other mobility models for ground users.}, as shown in Fig. \ref{environment}, where the blue cubes denote obstacles (e.g., buildings) with height $\text{40 m}$. It is assumed that the three mobile users follow the set trajectories to move, which cannot be known by the UAV prior to communication. For the considered practical segmented channel model, we use Rician fading to model the small-scale fading, and set $\beta_{k,\text{NLoS}}[n]=\text{--40 dB}$, $\beta_{k,\text{LoS}}[n]=\text{--30 dB}$, $\alpha_{k,\text{NLoS}}[n]=4$, $\alpha_{k,\text{LoS}}[n]=2$, $10\log_{10}(\xi_{k,\text{NLoS}}[n])\sim\tilde{\mathcal N}(0,5)$, and $10\log_{10}(\xi_{\text{k,LoS}}[n])\sim\tilde{\mathcal N}(0,2)$, where $\tilde{\mathcal N}$ denotes the Gaussian distribution. The UAV's initial location is set as $\hat{\mv q}_{I}=\text{(--95 m, --95 m)}$. To implement the table-based Q-learning, we divide the whole area into 20 by 20 grid, and set the UAV's displacement at each slot as $\lambda=\text{10 m}$. During the Q-learning, if the UAV's action leads to a location outside the 20 by 20 grid, we will add an additional negative reward. The learning rate is set as $\alpha=0.3$, the discount factor is set as $\gamma=0.9$ to make the UAV focus on long-term reward, and the probability for random actions $\epsilon$ is decreasing over time from the initial value of $\epsilon=0.9$. Furthermore, for the probabilistic LoS channel used in additional training procedure of enhanced RL-based approach,\footnote{In general, if we roughly know some statistical information about the areas (e.g., the intensity of buildings), we can choose channel parameters that are mostly suitable for this area. Otherwise, we choose  these parameters randomly.} we set $\bar{\alpha}=2.3$, $\bar{\beta}=\text{--30 dB}$, and $\eta=0.1$. For comparison, we consider the following maneuver design based on average channel power gains.

{\bf Heuristic maneuver design based on average channel power gains:} The UAV optimizes its action by assuming the channels to be probabilistic LoS channel. At slot $n\in\mathcal N$, the UAV knows the current user locations and solves the following sum-rate maximization problem:
\begin{align}
\text{(P2.$n$):}~\max\limits_{\bm q[n]}~&\log_{2}\left(1+\sum\limits_{k=1}^{K}\frac{P\bar{h}_{k}[n]}{\sigma^{2}}\right).\nonumber
\end{align}
The optimal solution to (P2.$n)$, denoted as $\tilde{\mv q}^{*}[n]$, can be obtained via a two-dimensional search in this area. Accordingly, the UAV chooses $\mv \lambda[n]\in\tilde{\mathcal A}$ such that it  moves towards $\tilde{\mv q}^{*}[n]$.

\begin{figure}
\centering
\includegraphics[width=7cm]{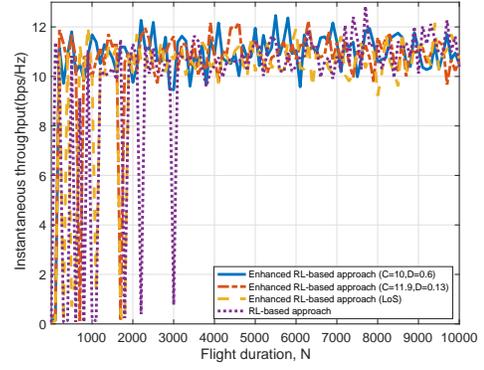}
\caption{Convergence behaviors of the proposed RL-based and enhanced RL-based approaches.}
\label{convergence}
\end{figure}
\begin{figure}
\centering
\includegraphics[width=7cm]{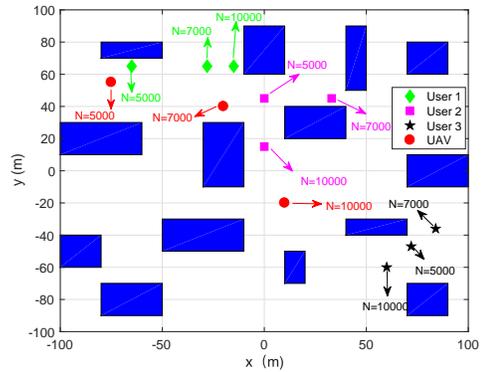}
\caption{UAV's locations after convergence at different particular time slots.}
\label{convergence_location}
\end{figure}

Fig. \ref{convergence} shows the instantaneous throughput of our proposed RL-based and enhanced RL-based approaches. First, it is observed that the proposed RL-based approaches almost converge at $N>4000$. Next, it is observed that for all proposed designs based on RL, at first, the UAV spend some time in exploring the unknown environment. Then, the UAV can track the ground users' movement and achieve the convergent performance. This is because as compared to the UAV's speed, the speed of ground user is slow under our simulation setup, such that the UAV is able to adjust its location to track the ground user's movement (see, Fig. \ref{convergence_location}). Finally, it is observed that enhanced RL-based designs have faster convergence speed as compared to the RL-based design and using the probabilistic LoS model for training procedure has faster convergence speed than the LoS channel. This indicates the significance of employing matching channel models for the enhanced RL-based designs in training procedure to speed up the convergence.

Fig. \ref{convergence_location}\footnote{Notice that in order to show that the UAV is able to track the users' movement, we choose some particular time slots to show the corresponding locations of UAV and ground users. In practice, the users' locations are changing over time, so that the UAV should move over time to provide better service to ground users.} shows the UAV's locations under the proposed RL-based designs at different particular time slots. It is observed that the UAV is able to track the ground users' movement, so as to achieve the convergent performance. This is consistent with the observation in Fig. \ref{convergence}. Besides, it is observed that the UAV can stay at the LoS propagations of most ground users to enhance the sum-rate throughput. This validates the effectiveness of our proposed RL-based designs.

\begin{figure}
\centering
\includegraphics[width=7cm]{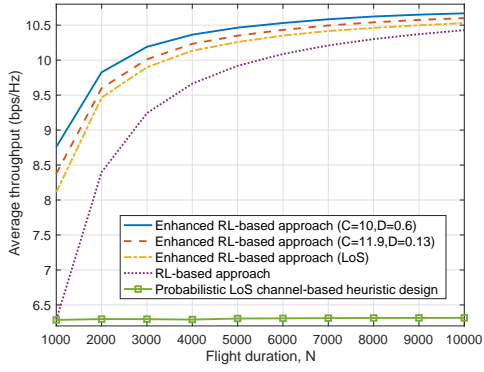}
\caption{Throughput comparison among RL-based, enhanced RL-based, and probabilistic LoS channel-based heuristic approaches.}
\label{rate_mobile_real}
\end{figure}
Fig. \ref{rate_mobile_real} shows the average throughput of our proposed designs. First, it is observed that RL-based designs significantly outperform the probabilistic LoS channel-based heuristic design. This is because the UAV-to-ground channels are NLoS with high probability due to the obstacles, which mismatch with the adopted probabilistic LoS channel model, thus leading to compromised performance for the probabilistic LoS channel-based heuristic approach. Next, the enhanced RL-based designs are observed to outperform the RL-based design, especially when the flight duration is small. This is because that the enhanced RL-based designs exploit the well-established wireless models as the expert knowledge to help initialize the Q-table to give rough information about the achievable rewards at different UAV locations, rather than setting all initial values of Q-table to zero as in the RL-based design. The performance gain indicates the significance of exploiting wireless expert knowledge to aid RL in speeding up the convergence. Finally, it is observed that the performance of  enhanced RL-based designs highly depends on the adopted channel models for training procedure. In particular, using the probabilistic LoS models is observed to achieve higher throughput than the LoS model. This demonstrates that employing a model that matches well with practical environment is essential for the enhanced RL-based designs.

\begin{figure}
\centering
\includegraphics[width=7cm]{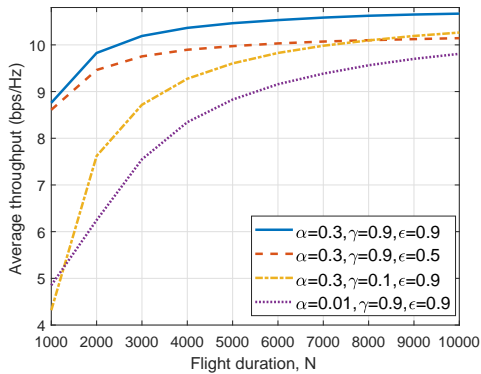}
\caption{Throughput comparison under different learning parameters.}
\label{rate_mobile_real2}
\end{figure}
Fig. \ref{rate_mobile_real2} shows the average throughput of the enhanced RL-based design under different learning parameters, with $C=10$ and $D=0.6$ for the employed probabilistic LoS channel model. In our setup, it is observed that using $\small{\alpha=0.3}$, $\small{\gamma=0.9}$, and $\small{\epsilon=0.9}$ achieves higher throughput than other parameters. This indicates the importance of choosing proper learning parameters in RL.

\section{Concluding Remarks}
This paper proposed the UAV maneuver design in real time for UAV-enabled uplink NOMA systems to maximize the sum-rate throughput of all users over a finite horizon, subject to practical constraints on the UAV's initial location and flight speed. We considered the more practical segmented channel model and assumed that  the UAV did not know any information about the channel environment and users' movement {\it a priori}. We proposed a novel approach to design the UAV maneuver based on RL via Q-learning, and building upon it we further proposed an enhanced RL-based approach, in which the wireless expert information on UAV-to-ground channel was utilized to facilitate the Q-learning via adding additional training procedure. Numerical results showed that our proposed RL-based and enhanced RL-based approaches significantly improved the sum-rate throughput, and the enhanced RL-based approach improved the convergence speed significantly as compared to the RL-based approach with zero Q-table initially.

Notice that in practice, the UAV can fly towards a variety of directions. However, increasing the number of possible actions will lead to higher model complexity. Therefore, in order to balance the tradeoff between the accuracy and model complexity, in this paper, we assume that the UAV flies at the fixed altitude similar as that in many prior works \cite{RZhangrelay, JXubroadcast,QWuscp}, and then only consider four directions for the UAV's movement. On possible extension to consider that the UAV can fly at any directions in 3D space instead of 4 horizontal directions. Under this setup, the state-action space in our proposed RL-based and enhanced RL-based approaches will become significantly huge, for which a deep Q-network based algorithm may necessary for efficiently solving this problem.

\end{document}